\documentclass{appolb}
\usepackage{amsthm}
\usepackage{amsmath}

\theoremstyle{plain}

\newcommand{\tmem}[1]{{\em #1\/}}
\newcommand{\op}[1]{#1}

\usepackage[english]{babel}

\begin{document}

\title{Effective Description of a Gauge Field and a Tower of Massive Vector
Resonances}

\author{Alfonso R. Zerwekh %
\thanks{alfonsozerwekh@uach.cl%
}
\address{ Centro de Estudios Subat\'omicos and Instituto de F\'isica,\\
  Facultad de Ciencias, Universidad Austral de Chile, \\
 Casilla 567, Valdivia, Chile}}
\maketitle
\begin{abstract}
In this work we review an effective description of the interaction
of a gauge field with a tower of massive vector fields by introducing
a non-diagonal mass matrix in a gauge invariant way. Particular cases
of the method with only one vector resonance have been used by the
author elsewhere, nevertheless in this paper the method is developed
in a general way and and we proof its main features for an arbitrary
number of vector resonances. Additionally, we show how to couple the
vector resonances with fermions. We find that the method can be useful
in order to describe the low energy phenomenology of scenarios like
Kaluza-Klein resonances of usual gauge bosons or Technicolor vector
resonances and detailed examples are provided. 
\end{abstract}


\section{Introduction}

There are some physical scenarios where gauge bosons interact and
mix with massive vector bosons. Perhaps the best known example is
the case of the photon, the rho and the omega (if isospin violation
is allowed) in hadron physics. In fact, this is the original motivation
for the development of the Vector Meson Dominance Hypothesis (VMD)(for
a review see \cite{vmd}). On the other hand, some extensions of the
Standard Model include massive vector fields which mix with vector
gauge bosons. In Technicolor models, for example, we find the technirho
and the techniomega, which mix with the photon and the Z, and the
color octet technirho, which mixes with the gluon. In Topcolor assisted
Technicolor models the gluon mixes with four color octet technirhos
and the so called colorons (see, for example \cite{Lane_Mrenna}).
Another important case is the interaction of the usual gauge bosons
(for example the gluon) with their Kaluza-Klein resonances in models
with extra dimensions. Another recent example is Higgsless model.
The expectation of finding these resonances at the LHC has strongly
motivated the question about how to describe them in the context of
a four-dimensional phenomenological model\cite{Hill}. For example,
the so called Deconstruction Theory was first invented to describe
the Kaluza-Klein resonance of the gluon \cite{Hill,Decon}. In this
kind of models, a set of {}``hidden local symmetries'' is introduced.
The breaking down of the hidden symmetry is described by a non-linear
sigma model and the massive gauge fields associated with the {}``hidden
symmetries'' are identified with the Kaluza-Klein resonances.

In this paper we review a general method for describing the interaction
of a gauge field with a set of massive vector resonances based on
the observed low energy local symmetries. This method can be useful
as a first effective approach for studying new vector resonances.
Particular cases of this method have been used for studying the phenomenology
of the color-octet technirho \cite{Zerwekh:2006te} and the axigluon
\cite{Zerwekh:2009vi}. We build the method in a general way and we
proof its more important property for an arbitrary number of resonances.
Additionally, important particular cases are studied.

We have organized this work as follow. In section 2 we present the
basic Lagrangian and its properties (including the coupling of the
resonances with fermions). Section 3 is devoted to show some examples
and applications. Finally, some conclusions are presented.

\section{The Lagrangian}

In order to fix ideas, let us consider an extension of QCD defined
by a Lagrangian of the form: \begin{equation}
\mathcal{L}=-\frac{1}{4}\sum_{i}\tilde{F}_{i\mu\nu}^{a}\tilde{F}_{i}^{a\mu\nu}+\sum_{i\, j}\frac{1}{2}A_{i\mu}^{a}\mathcal{M}_{ij}^{2}A_{j}^{a\mu}\label{lagrangian}\end{equation}
 with \begin{equation}
\tilde{F}_{i\mu\nu}^{a}=\partial_{\mu}A_{i\nu}^{a}-\partial_{\nu}A_{i\mu}^{a}-\tilde{g}_{i}f^{abc}A_{i\mu}^{b}A_{i\nu}^{c}\end{equation}
 where $a$ is a group index and $i=1\ldots N$ is the {}``flavor''
of the vector field. In order to make the notation clear we have written
explicitly the sum over the flavor indices. We assume that all the
vector fields transform under the \textit{same} group {\tmem{SU(3)}}
as follow: \begin{equation}
\delta A_{i}^{a}=-f^{\op{abc}}A_{i}^{b}\Lambda^{c}-\frac{1}{\tilde{g}_{i}}\partial\Lambda^{a}\label{Atransform}\end{equation}
 From equations (\ref{lagrangian}) and (\ref{Atransform}) we find
that the variation of the Lagrangian is : \begin{equation}
\delta\mathcal{L}=\sum_{i\, j}A_{i\mu}^{a}\mathcal{M}_{ij}^{2}\frac{1}{\tilde{g}_{j}}\partial^{\mu}\Lambda^{a}\end{equation}
 Evidently, $\delta\mathcal{L}$ must vanish independently of the
fields, so the following equation must be satisfied: \begin{equation}
\sum_{j}\mathcal{M}_{ij}^{2}\frac{1}{\tilde{g}_{j}}=0\label{eigeneq}\end{equation}
 We can solve equation (\ref{eigeneq}) for the diagonal term and
we obtain : \begin{equation}
M_{i}^{2}\equiv\mathcal{M}_{ii}^{2}=-\tilde{g}_{i}\sum_{j\neq i}\mathcal{M}_{ij}^{2}\frac{1}{\tilde{g}_{j}}\label{masseq}\end{equation}

\subsection{Transformation of the Physical Fields}

Equation (\ref{eigeneq}) shows us that the vector \begin{equation}
v_{1}=\frac{1}{\sqrt{\sum_{i}\frac{1}{\tilde{g}_{i}^{2}}}}\left(\begin{array}{c}
\frac{1}{\tilde{g}_{1}}\\
.\\
.\\
.\\
\frac{1}{\tilde{g}_{N}}\end{array}\right)\label{v1}\end{equation}
 is an eigenvector of $\mathcal{M}^{2}$ with null eigenvalue. We
want to emphasize that this is the necessary and sufficient condition
to include a mass matrix for the gauge sector of a gauge invariant
theory.

Let us now define a new field:

\begin{equation}
G_{1}=g\sum_{i}\frac{A_{i}}{\tilde{g}_{i}}\label{G1}\end{equation}
 where $g$ is defined by \begin{equation}
\frac{1}{g^{2}}\equiv\sum_{i}\frac{1}{\tilde{g}_{i}^{2}}\label{defg}\end{equation}
 It is clear that, by construction, $G_{1}$ is massless. Using equation
(\ref{Atransform}) and (\ref{G1}) it is easy to see that $G_{1}$transforms
in the following way: \begin{equation}
\delta G_{1}^{a}=-f^{\op{abc}}G_{1}^{b}\Lambda^{c}-\frac{1}{g}\partial\Lambda^{a}\label{G1transform}\end{equation}
 That is, $G_{1}$ transforms as a true gauge field.

Because matrix $\mathcal{M}^{2}$ is hermitian, all other independent
eigenvectors must be orthogonal to (\ref{v1}). Let us call such vectors
as $v_{i}$ ($i=2\ldots N)$. We can now define: \begin{equation}
G_{i}=\sum_{j}v_{ij}A_{j}\label{Gi}\end{equation}
 with $i=2\ldots N$, $j=1\ldots N$ and $v_{ij}$ such that \begin{equation}
\sum_{j}v_{ij}\frac{1}{\tilde{g}_{j}}=0\label{ortho}\end{equation}
 Equation (\ref{ortho}) represents the orthogonality condition with
$G_{1}$. Using equation (\ref{Atransform}), we find that $G_{i}$
transforms as: \begin{equation}
\delta G_{i}^{a}=-\sum_{j}v_{ij}f^{\op{abc}}A_{j}^{b}\Lambda^{c}-\sum_{j}v_{ij}\frac{1}{\tilde{g}_{j}}\partial\Lambda^{a}\end{equation}
 From (\ref{Gi}) and (\ref{ortho}) we obtain: \begin{equation}
\delta G_{i}^{a}=-f^{\op{abc}}G_{i}^{b}\Lambda^{c}\mbox{\hspace{1cm}with }i=2\ldots N\label{Gitransform}\end{equation}
 That means that all the all the massive resonances $G_{i}$ ($i\neq1$)
transform like matter fields in the adjoint representation.

\subsection{Couplings with Fermions}

Any realistic model must consider the coupling of vector fields with
fermions. In gauge theories, this coupling is obtained in a minimal
way by defining a covariant derivative. Nevertheless, in our case,
we have too many vector fields transforming in a gauge-like way under
the {\tmem{same}} group. We can, then, ask whether it is possible
to build a gauge invariant Lagrangian that describe the interaction
of fermions with all these vectors fields. We propose the following
simple solution. Let us consider the Lagrangian: \begin{equation}
\mathcal{L}=\bar{\psi}(i\gamma^{\mu}D_{\mu}-m)\psi\label{lfA1}\end{equation}
 With $D_{\mu}$ given by: \begin{equation}
D_{\mu}\equiv\partial_{\mu}+i\frac{\lambda^{a}}{2}\sum_{j}\tilde{g}_{j}f_{j}A_{j\mu}^{a}\label{lfA2}\end{equation}
 and \begin{equation}
\sum_{i}f_{i}=1\label{lfA3}\end{equation}
 It is clear that the variation of each $A_{j\mu}^{a}$ cancels a
fraction $f_{j}$ of the total variation coming from the kinetic term
of the fermions. Equation (\ref{lfA3}) ensures the complete gauge
invariance of the Lagrangian%
\footnote{Perhaps a more rigorous but equivalent way to obtain our generalized
covariant derivative is defining first an usual covariant derivative
for each gauge-like field:\[
D_{j\mu}=\partial_{\mu}+i\frac{\lambda^{a}}{2}\tilde{g}_{j}A_{j\mu}^{a}\]

We can, then, construct the covariant derivative as a weighted sum:

\[
D_{\mu}=\sum_{j}f_{j}D_{j\mu}\]

where the set of weights $\{f_{j}\}$ satisfy equation (\ref{lfA3})
and can be viewed as a normalized discrete distribution. On the other
hand, we can use $D_{j\mu}$ to define formally the field strength
tensor for each gauge-like field:

\[
i\tilde{g}_{j}\frac{\lambda^{a}}{2}\tilde{F}_{j\mu\nu}^{a}\equiv\left[D_{j\mu},D_{j\nu}\right]\]
}.

It is interesting to investigate how the field $G_{1}$ couples to
fermions. The first step is to realize that, after the change of basis,
Lagrangian (\ref{lfA1}) takes the form: \begin{equation}
\mathcal{L}=\mathcal{L}_{0}+\mathcal{L}_{1}+\mathcal{L}_{2}\label{lfG1}\end{equation}
 where: \begin{equation}
\mathcal{L}_{0}=\bar{\psi}(i\gamma^{\mu}\partial_{\mu}-m)\psi\label{lfG2}\end{equation}
 \begin{equation}
\mathcal{L}_{1}=-\xi_{1}\bar{\psi}\frac{\lambda^{a}}{2}\gamma^{\mu}G_{1\mu}^{a}\psi\end{equation}
 \begin{equation}
\mathcal{L}_{2}=-\xi_{j}\bar{\psi}\frac{\lambda^{a}}{2}\gamma^{\mu}G_{j\mu}^{a}\psi\mbox{\hspace{1cm}( }j=2\ldots N\mbox{ )}\label{lfG3}\end{equation}
 with $\xi_{i}$ ($i=1\ldots N)$ are some coupling constants. It
is easy to see, by equation (\ref{Gitransform}), that $\delta\mathcal{L}_{2}=0$.
That means that, in order to have a gauge invariant Lagrangian, $\delta\mathcal{L}_{1}$
must cancel the variation of the fermionic kinetic term. However,
in virtue of (\ref{G1transform}), this cancellation only happens
if $\xi_{1}=g$. That is, gauge invariance warrants that $G_{1}$
couple to fermions as a true gauge boson independently of the value
of $f_{i}$ constants.

Conversely, we can start from the Lagrangian defined by equations
(\ref{lfG1})-(\ref{lfG3}) (with $\xi_{1}=g$) and rotate the fields
$G_{i}$ by using equations (\ref{G1}) and (\ref{Gi}). The resulting
Lagrangian can be written as: \begin{equation}
\mathcal{L}_{}=\bar{\psi}(i\gamma^{\mu}\partial_{\mu}-m)\psi-\bar{\psi}\frac{\lambda^{a}}{2}\gamma^{\mu}\psi\sum_{i}\left[\frac{g}{\tilde{g}_{i}}+\sum_{j}\xi_{j}v_{ji}\right]A_{i\mu}^{a}\end{equation}
 If we make the definition: \begin{equation}
f_{i}\equiv\frac{1}{\tilde{g}_{i}}\left[\frac{g}{\tilde{g}_{i}}+\sum_{j}\xi_{j}v_{ji}\right]\label{deffi}\end{equation}
 (without sum in the repeated indexes) then we obtain exactly the
Lagrangian defined by equations (\ref{lfA1}) and (\ref{lfA2}). On
the other hand, if we sum all the $f_{i}$ defined in (\ref{deffi})
and we use equations (\ref{defg}) and (\ref{ortho}) then we obtain
the condition $\sum_{i}f_{i}=1$.

\section{Examples and Applications}

\subsection{Case $N=2$}

The gauge sector of this case was partially studied in a previous
work \cite{AZ_RR} on the gluon-color octet technirho mixing {%
\footnote{In this case, however, additional dimension six terms must be taken
into account in order to describe effects due to the technirho compositeness
\cite{Chiv_Sim}.%
}}. Nevertheless it is worth to review this important and simple case
in order to illustrated the methods described here.

In this case the Lagrangian of the gauge sector can be written as:

\begin{equation}
\mathcal{L}=-\frac{1}{4}\tilde{F}_{1\mu\nu}^{a}\tilde{F}_{1}^{a\mu\nu}-\frac{1}{4}\tilde{F}_{2\mu\nu}^{a}\tilde{F}_{2}^{a\mu\nu}+\frac{M^{2}}{2\tilde{g}_{2}}\left(\tilde{g}_{1}A_{1\mu}^{a}-\tilde{g}_{2}A_{2\mu}^{a}\right)^{2}\label{eq:N2}\end{equation}

where $M$ is a new mass scale present in the model. We want to emphasize
this important point. It is an intrinsic feature of the class of model
studied in this work the appearance of new mass scales that are compatible
with gauge symmetry but remains unconstrained by it.

The mass matrix originated by this Lagrangian (which is in this case
the most general mass matrix for the gauge sector compatible with
gauge symmetry) is exactly diagonalizable. In fact, the fields which
are mass eigenstates can be written as

\begin{eqnarray}
G_{1} & = & A_{1}\cos(\alpha)+A_{2}\sin(\alpha)\label{eq:EigenVector1}\\
G_{2} & = & -A_{1}\sin(\alpha)+A_{2}\cos(\alpha)\label{eq:EigenVector2}\end{eqnarray}

where $\sin(\alpha)=g/\tilde{g}_{2}$ and $g$ is, according to equation
(\ref{defg}),

\[
g=\frac{\tilde{g}_{1}\tilde{g_{2}}}{\sqrt{\tilde{g}_{1}^{2}+\tilde{g}_{2}^{2}}}.\]

The mass of the physical states are:

\begin{eqnarray*}
m_{G_{1}} & = & 0\\
m_{G_{2}} & = & \frac{M}{cos(\alpha)}\end{eqnarray*}

so, we see that the Lagrangian describes, in fact, a massless gauge
boson and a massive spin-one resonance in the adjoint representation.

Let us study, for a moment, the decay process $G_{2}\rightarrow G_{1}G_{1}$.
The relevant part of the Lagrangian, written in terms of the physical
fields, is:

\begin{eqnarray}
\mathcal{L}_{G_{2}G_{1}G_{1}} & = & \left(-g_{1}\cos^{2}(\alpha)\sin(\alpha)+g_{2}\cos(\alpha)\sin^{2}(\alpha)\right)f^{abc}\{\partial_{\mu}G_{1\nu}^{a}G_{1}^{b\mu}G_{2}^{c\nu}+\notag\\
 &  & +\partial_{\mu}G_{1\nu}^{a}G_{2}^{b\mu}G_{1}^{c\nu}+\partial_{\mu}G_{2\nu}^{a}G_{1}^{b\mu}G_{1}^{c\nu}\}\label{eq:L211}\end{eqnarray}

It is clear that, due to the definition of $\alpha$, $\tilde{g}_{1}\cos(\alpha)=\tilde{g}_{2}\sin(\alpha)$.
Using this identity, we can see that the coupling constant of the
$G_{2}G_{1}G_{1}$interactions vanishes exactly. This result is an
important consequence of the gauge symmetry of the model.

Let us turn our attention to the coupling with fermions. When we write
Lagrangian (\ref{lfA1}) in terms of physical fields we obtain: \begin{equation}
\mathcal{L}=\bar{\psi}\left[i\gamma^{\mu}\partial_{\mu}-m-g\gamma^{\mu}G_{1\mu}^{a}\frac{\lambda^{a}}{2}-\frac{(\tilde{g}_{2}^{2}f_{2}-\tilde{g}_{1}^{2}f_{1})}{\sqrt{\tilde{g}_{1}^{2}+\tilde{g}_{2}^{2}}}\gamma^{\mu}G_{2\mu}^{a}\frac{\lambda^{a}}{2}\right]\psi\label{lN=2}\end{equation}
 Two interesting cases arise: 
\begin{enumerate}
\item We can impose that $\tilde{g}_{1}=\tilde{g}_{2}$ and $f_{1}=f_{2}$.
In this case, $G_{2}$ completely decouple from the fermionic sector.
On the other hand, it is not difficult to see that $G_{2}$ interacts
with the gauge boson ($G_{1}$) only through terms of the form $G_{1}G_{2}G_{2}$
and $G_{1}G_{1}G_{2}G_{2}$ (In the case of the technirho other high
dimensional operators must be taken into account due to its compositeness
nature \cite{Chiv_Sim}, but here we can assume that $G_{2}$ is an
elementary field). All that mean that a symmetric choice of the constants
makes $G_{2}$ to be stable. 
\item Another interesting possibility come to us when we realize that the
constants $f_{1}$ and $f_{2}$ can be different for different fermionic
generations. We can for example choose that constants in such a way
that $G_{2}$ decouple from the first two generations but couple strongly
to the third one. In this way it may be possible to build an effective
Topcolor model. 
\end{enumerate}
Finally, we must remark that Lagrangian (\ref{lN=2}) is a generalization
of VMD. In fact, we obtain the traditional VMD result by choosing
$f_{2}=0$ and $f_{1}=1$%
\footnote{The model studied in \cite{AZ_RR} is based on the VMD hypothesis.
That means that the vector resonance (the proto-technirho) couple
to the quarks only through the mixing with the proto-gluon. Nevertheless
Extended Technicolor must produce a direct (proto-)technirho-quark
interaction. The formalism presented here allows us to include such
interactions %
}.

\subsection{Case $N=2$ with Symmetry Breaking}

Let us consider again Lagrangian (\ref{eq:N2}) but now we add a non-linear
sigma model term:

\begin{equation}
v^{2}\mathrm{tr}\left[D_{\mu}U^{\dagger}D^{\mu}U\right]\end{equation}

where $v$ is some energy scale at which the gauge symmetry spontaneously
breaks down, and $D_{\mu}U$ is:

\begin{equation}
D_{\mu}U=\partial_{\mu}U-\frac{i}{2}\tilde{g}_{1}A_{1\mu}U-\frac{i}{2}\tilde{g}_{2}A_{2\mu}U\end{equation}

Notice that we have put in the covariant derivative $A_{1}$ as well
as $A_{2}$ with equal relative weight. In the unitary gauge ($\left\langle U\right\rangle =1$),
the non-linear sigma model term produce additional contributions to
the mass matrix. The resulting mass matrix can be exactly diagonalized
but, as usual, in order to simplify our expressions we will consider
only the case where $\tilde{g}_{2}\gg\tilde{g}_{1}$ . In this limits
the mass eigenvalues are:

\begin{eqnarray*}
m_{1}^{2} & = & \frac{2\tilde{g}_{1}v^{2}}{1+\frac{\tilde{g}_{2}v^{2}}{2M^{2}}}\\
m_{2}^{2} & = & M^{2}\left(1+\frac{\tilde{g}_{2}v^{2}}{2M^{2}}\right)\end{eqnarray*}

Of course, the physical fields can be written again in the form given
in equations (\ref{eq:EigenVector1}) and (\ref{eq:EigenVector2}),
but now the mixing angle is given by:

\begin{equation}
\tan(\alpha)=\frac{\tilde{g_{1}}}{\tilde{g_{2}}}\left\{ \frac{1-\frac{\tilde{g}_{2}v^{2}}{2M^{2}}}{1+\frac{\tilde{g}_{2}v^{2}}{2M^{2}}}\right\} \label{eq:tana}\end{equation}

Now we will consider again the decay $G_{2}\rightarrow G_{1}G_{1}$.
This process is still described by Lagrangian (\ref{eq:N2}) but this
time $\tilde{g}_{1}\cos(\alpha)\neq\tilde{g}_{2}\sin(\alpha)$ and
the coupling constant of this process doesn't vanish. In fact, using
the expression for $\tan(\alpha)$ we wrote above, we can see that
the Lagrangian (\ref{eq:N2}) can be written as:

\begin{eqnarray}
\mathcal{L}_{G_{2}G_{1}G_{1}} & =-\frac{2\tilde{g}_{1}^{2}v^{2}}{2M^{2}+\tilde{g}_{2}v^{2}}\left(\frac{1-\frac{\tilde{g}_{2}v^{2}}{2M^{2}}}{1+\frac{\tilde{g}_{2}v^{2}}{2M^{2}}}\right)^{2} & f^{abc}\{\partial_{\mu}G_{1\nu}^{a}G_{1}^{b\mu}G_{2}^{c\nu}+\notag\\
 &  & +\partial_{\mu}G_{1\nu}^{a}G_{2}^{b\mu}G_{1}^{c\nu}+\partial_{\mu}G_{2\nu}^{a}G_{1}^{b\mu}G_{1}^{c\nu}\}\end{eqnarray}

A similar phenomenon occurs when we couple the vector bosons to fermions.
As we did above, we implement the minimal coupling defining a generalized
covariant derivative. Thus, we write:

\begin{equation}
\mathcal{L}=\bar{\psi}(i\gamma^{\mu}D_{\mu}-m)\psi\end{equation}

with 

\begin{equation}
D_{\mu}=\partial_{\mu}+i\frac{\lambda^{a}}{2}(1-f)\tilde{g}_{1}A_{1\mu}^{2}+i\frac{\lambda^{a}}{2}f\tilde{g}_{2}A_{2\mu}^{2}\end{equation}

When we make the rotation to the physical basis, the interaction term
between the fermions and the {}``would-be'' gauge boson can be written
as:

\begin{equation}
\left[\tilde{g}_{1}\cos(\alpha)+f\left(\tilde{g}_{2}\sin(\alpha)-\tilde{g}_{1}\cos(\alpha)\right)\right]\bar{\psi}G_{1\mu}^{a}\frac{\lambda^{a}}{2}\gamma^{\mu}\psi\end{equation}

Again, because of equation (\ref{eq:tana}), $\tilde{g}_{2}\sin(\alpha)-\tilde{g}_{1}\cos(\alpha)\neq0$.
This means that, due to the symmetry breaking the coupling constant
characteristic of the interaction between the fermions and the {}``would-be''
gauge boson, depends on the free parameter $f$.

\subsection{First Neighbor Coupling on a Ring}

Another interesting case is to study $N$ vector bosons with first
neighbor coupling on a ring. In this case, equation (\ref{masseq})
reduce to: \begin{equation}
M_{i}^{2}=-\frac{\tilde{g}_{i}}{\tilde{g}_{i-1}}\mathcal{M}_{ii-1}^{2}-\frac{\tilde{g}_{i}}{\tilde{g}_{i+1}}\mathcal{M}_{ii+1}^{2}\end{equation}
 (without sum in the repeated indexes) and we impose periodic boundary
 conditions:
 \begin{equation}
\mathcal{M}_{10}^{2}=\mathcal{M}_{1N}^{2}\end{equation}
 \begin{equation}
\mathcal{M}_{NN+1}^{2}=\mathcal{M}_{N1}^{2}\end{equation}
 If additionally we assume that all the coupling constants are identical
and $\mathcal{M}_{ii-1}^{2}=\mathcal{M}_{ii+1}^{2_{}}=k^{2}$ (with
$k^{2}$ some arbitrary constant) we get: \begin{equation}
M_{i}^{2}=-2k^{2}\end{equation}
 Finally, the bosonic mass matrix has the form: \begin{equation}
\mathcal{M}^{2}=k^{2}\left(\begin{array}{ccccccccc}
-2 & 1 & 0 & 0 & . & . & . & 0 & 1\\
1 & -2 & 1 & 0 & . & . & . & 0 & 0\\
0 & 1 & .\\
. & 0 &  & .\\
. & . &  &  & .\\
. & . &  &  &  & .\\
. & . &  &  &  &  & .\\
0 & . &  &  &  &  &  & .\\
1 & 0 &  &  &  &  &  &  & -2\end{array}\right)\end{equation}

It is well known that the eigenvalues of this mass matrix describe
(for $N\gg1$) a Kaluza-Klein like spectrum.

\section{Conclusions}

In this work we have studied a general method for describing the mixing
of an arbitrary number of massive vector bosons with a gauge boson,
and their interactions.The key idea is the introduction of a
non-diagonal mass matrix in the gauge sector in a gauge invariant
way. The coupling of all the vector bosons to fermions
has been, also, treated in a consistent way. The resulting Lagrangian represents
a generalization of the old idea of Vector Meson Dominance. We have
shown that this method can be useful as a phenomenological description
of a Kaluza-Klein-like tower of vector resonances. Other scenarios
like Topcolor model can be also described.

Of course, the use of the method presented here aims to be only a
low energy phenomenological description of the physics of vector resonances,
but we think it can serve as a useful first approach if they are discovered
at the LHC

\section*{Acknowledgments}

The author wishes to thank Claudio Dib and Rogerio Rosenfeld for interesting
and useful discussions. This work was supported by Fondecyt (Chile)
grant No. 1070880.TGD

\end{document}